\documentclass{revtex4-2}
\usepackage{amsmath, amssymb}
\usepackage{graphicx}
\usepackage{url}
\usepackage{hyperref}
\usepackage{appendix}
\usepackage{float}
\usepackage{booktabs}
\usepackage{tabularx}

\def\be{\begin{equation}}
	\def\ee{\end{equation}}
\def\bea{\begin{eqnarray}}
	\def\eea{\end{eqnarray}}

\begin{document}
	
	\title{Unifying Cosmic Epochs and Effective Dynamical Brane Tension}
	
	\author{Farzin Safarzadeh-Maleki}
	\affiliation{School of Astronomy, Institute for Research in Fundamental Sciences (IPM), P.O. Box 19395-5531, Tehran, Iran}
	\email{f.safarzadeh@ipm.ir}

	\
	\
	
	\begin{abstract}

	We present a single analytic scale factor \(a(t)=e^{H(t)t}\bigl(1-e^{-k(t)t}\bigr)^{b(t)}\) that unifies the cosmic expansion history from inflation to the present in a smooth and differentiable manner. By eliminating conventional piecewise definitions, the model provides a compact global description of cosmic evolution while remaining consistent with SNIa, CMB, and BAO observations.

The model introduces a time-dependent scale-modulation parameter $k(t)$, with the dimensionless combination $x(t)$ governing the asymptotic evolution across all epochs. This construction finds a compelling realization as a brane-inspired effective framework: 
interpreted as the ratio $x(t)\sim\rho(t)/\lambda(t)$ of the energy density to a dynamical brane tension, the model necessitates a time-varying $\lambda(t)$, thereby driving acceleration without a cosmological constant. This single parameter bridges the observational successes of the Randall-Sundrum II and chrono-brane paradigms within a coherent, effective framework. 
Although phenomenological in origin, the construction offers an analytically tractable and observationally testable alternative to $\Lambda$CDM that requires neither a fixed vacuum energy nor abrupt phase transitions.
		
	\end{abstract}
	
	
	\maketitle
	
	\section{Introduction}
	\label{sec:intro}
	
		A central goal of modern theoretical cosmology is to understand the universe's expansion history in a unified manner. 
		The standard description of the universe’s expansion history is built from piecewise solutions: exponential growth during inflation, power‑law behaviour during radiation and matter domination, and exponential acceleration at late times~\cite{Weinberg, Baumann, Mukhanov}. 
		Although each cosmological epoch admits a complete local description, transitions between epochs are often modeled by patching together distinct functional forms of the scale factor, yielding a piecewise description that may exhibit non-smooth behavior and thereby complicate both theoretical modeling and numerical simulations.
		Many attempts have been made to unify these epochs. Quintessential inflation employs a single scalar field to drive both early and late-time acceleration~\cite{Dimopoulos2002, Sami2004, deHaro2016}; modified gravity theories (e.g. \(f(R)\)) also underscore the desire to describe early time inflation and late time acceleration~\cite{Nojiri2003, Nojiri2017}; and various parametrisations of the Hubble rate or scale factor have been proposed~\cite{Aydiner2022, Singh2021}. 
	 	 Nevertheless, a simple, minimal analytic formula, one that smoothly interpolates across all four cosmic epochs while retaining a direct and transparent physical interpretation, has remained elusive. Such a single, continuous expression spanning the entire expansion history would serve not only as a practical computational tool but also as a suggestive hint toward a deeper unifying principle.
	 	 
		In this work we propose a unified scale factor $a(t)=e^{H(t)\,t}\Bigl(1-e^{-k(t)t}\Bigr)^{b(t)} $ as a phenomenological framework that interpolates between the four standard cosmological epochs and recovers the correct asymptotic behaviors in each era. 
	 To capture the full cosmic evolution within a single continuous framework, we reconstruct the key parameters of scale factor using smooth sigmoid functions ~\cite{Danca2015, Martinelli2023} that ensure both monotonic growth and continuous differentiability across the transition points. This construction 
	 is guided by physically motivated characteristic timescales and a consistent number of e-folds, yielding an evolution that accurately traces the expansion history from inflation to the present day. The resulting global and zoomed evolution exhibits agreement with SNIa, CMB, and BAO observations~\cite{Riess}-\cite{Eisenstein}.

	We demonstrate that a single time-dependent parameter of the model can capture the entire cosmic history without invoking separate mechanisms or a cosmological constant. We introduce a scale-modulation function $k(t)$ that controls the saturation rate of the exponential contributions in the evolution function. The associated dimensionless combination $x(t)=k(t)t$ then governs the full asymptotic evolution, smoothly steering the universe through inflation, the radiation and matter-dominated eras, and late-time acceleration without recourse to 
	artificial phase splicing.

	The unified scale-factor ansatz presented here finds a compelling realization as a brane-inspired effective framework: Physically, $x(t)$ admits a direct interpretation as the ratio $\rho(t)/\lambda(t)$ of the total energy density to a time-dependent brane tension. Allowing $\lambda(t)$ to vary with cosmic time, such correspondence reproduces the full expansion history without a cosmological constant. This dynamical behavior enables a reconstruction of $\lambda(t)$ from the scale factor $a(t)$, encoding the entire cosmic trajectory as a single brane evolution with evolving tension, wherein $x(t)$ naturally acts as a screening function that suppresses bulk corrections precisely when the energy density is low ($\rho \ll \lambda$). Although inherently phenomenological, the construction establishes a coherent effective framework  
	anchored in Randall-Sundrum type-II~\cite{Randall1999a,Randall1999b} 
	and its variable-tension extensions~\cite{GarciaAspeitia2018a}. 
	This same parameter effectively bridges the observational successes of the Randall-Sundrum II and chrono-brane paradigms across all epochs, offering a compelling phenomenological alternative to $\Lambda$CDM that requires neither a fixed cosmological constant nor artificial phase transitions. 
	
	The construction is analytically tractable, observationally testable, and provides a fresh perspective on the unification of cosmic epochs. 
	It suggests a dynamical reinterpretation of the effective vacuum energy: the onset of acceleration emerges not as an isolated late-time add-on, but as a natural consequence of the same unifying principle that determines the inflationary and intermediate epochs. Furthermore, the mathematical structure relating the early and late-time de Sitter asymptotes hints at a possible deeper organizing principle, the investigation of which we outline as a promising direction for future work.

Throughout the paper, we formulate our reconstruction directly in cosmic time $t$, rather than the more observationally familiar redshift $z$. Although the formal mapping $	t(z) = \int_{z}^{\infty} \frac{\mathrm{d}z'}{(1+z') H(z')}$ is mathematically exact, it is not observationally reconstructible, since its evaluation requires knowledge of the expansion history at arbitrarily high redshift, including the reheating epoch, whose dynamics remain unconstrained by direct observations.
 Furthermore, our reconstructed brane tension $\lambda$ is prescribed as a function of cosmic time itself, so that recasting it into the redshift frame would artificially require the same non-local inversion. By retaining $t$ as our primary variable, we preserve the manifestly local nature of the Friedmann equations, avoid any implicit dependence on reheating physics, and maintain a parametrization that is theoretically self-contained.
We emphasize that this is a choice of theoretical clarity. 
Redshift remains the natural variable for observations, and we provide the corresponding ranges for orientation: inflation ($z \gtrsim 10^{28}$), radiation ($3400 \lesssim z \lesssim 10^{28}$), matter ($1 \lesssim z \lesssim 3400$), and late-time acceleration ($z \lesssim 1$).

	The paper is organised as follows. Section~\ref{sec:model} introduces the scale factor and its piecewise definition, and demonstrates how the asymptotic limits recover the standard epochs. Section~\ref{sec:smooth} describes the sigmoid smoothing that makes the model fully continuous and differentiable. Section~\ref{sec:brane} presents the brane‑world interpretation, including the role of \(x(t)\) as \(\rho/\lambda\) and the emergence of a dynamical brane tension. Section~\ref{sec:Discussion} points to several avenues for future exploration. We conclude in Section~\ref{sec:conclusion} with a summary of our findings and their implications for a unified description of cosmic evolution.

	\section{A unified scale factor}
	\label{sec:model}
	
	We propose the following analytic form for the scale factor
	\begin{equation}
		a(t)=e^{H(t)\,t}\Bigl(1-e^{-k(t)t}\Bigr)^{b(t)} \label{eq:scale}
	\end{equation}
	where \(H(t)\), \(k(t)\), and \(b(t)\) are  functions of cosmic time \(t\). 
	In the exponential factor \(e^{H(t)t}\), which is dominated during inflation and at late times, the dynamical evolving parameter $H(t)$ is given by
		\bea
	H(t) =
	\begin{cases}
		H_{\mathrm{inf}} =  10^{37} \, & 10^{-36} < t \leq 10^{-32} \quad \text{(inflation)} \\[8pt]
		H_{\mathrm{inter}} = \dfrac{b(t)}{t} \,  & 10^{-32} < t < 10^{17} \quad \text{(intermediate)}\\[8pt]
		H_{\mathrm{DE}} = 2.2 \times 10^{-18} \,  & t \geq 10^{17} \quad \text{(dark energy)}
	\end{cases}\label{eq:H(t)}
	\eea
	  This parameter controls the exponential expansion and encodes dynamical scaling behavior that resembles Hubble-like evolution. The sub-indices "inf", "inter", and "DE" denote inflation, intermediate stages and dark energy era respectively, with values based on Planck and BAO data \cite {Aghanim}. Note that all quantities are in units where $[H] = \mathrm{s}^{-1}$.
	The second factor in the ansatz, $ { (1-e^{-k(t)t}) }^{b(t)}$, operates as a smoothing or saturation component. It becomes dominant at intermediate times, mediating the transition between radiation and matter-dominated regimes while preserving their expected expansion behavior. Within this framework, the exponent definition 
	\bea 
	b(t) = \frac{1}{2} \frac{1}{1 + \frac{t}{t_{\mathrm{eq}}}} + \frac{2}{3} \frac{\frac{t}{t_{\mathrm{eq}}}}{1 + \frac{t}{t_{\mathrm{eq}}}},\label{eq:b(t)}
	\eea
	with radiation-matter equality time $t_{\mathrm{eq}} \approx 10^{12}$ s, provides a smooth interpolation between \(b=1/2\) (radiation domination, \(t\ll t_{\mathrm{eq}}\)) and \(b=2/3\) (matter domination, \(t\gg t_{\mathrm{eq}}\)) and permits the scale factor to follow the expected power-law evolution in each regime.  
	Furthermore, the parameter $k(t)$ plays the role of an effective transition rate, setting the characteristic timescale over which the evolution shifts from power-law to exponential behaviour. It is defined as
	\bea
	k(t) =
	\begin{cases}
		k_{\mathrm{inf}} \geq 10^{39}, & 10^{-36} < t \leq 10^{-32} \\[8pt]
		k_{\mathrm{inter}} = \frac{1}{e\, b(t)} \sqrt{\frac{8 \pi G}{3} \rho_0^{\mathrm{RD,MD}}}, & 10^{-32} < t < 10^{17} \\[8pt]
		k_{\mathrm{DE}} \geq 10^{-14}, & t \geq 10^{17}
	\end{cases} \label{eq:k(t)}
	\eea
	in natural units such that $[k]=\mathrm{s}^{-1}$ and serves as a scale modulating parameter in the energy evolution function. In this relation the values associated with inflationary and dark energy phases are not chosen arbitrarily. Rather, the structure of the model imposes lower bounds on these parameters to ensure $k(t)t\gg1$, thereby securing the required regime of exponential expansion. 
	 During the intermediate epoch,  $\rho_{0}^{(\mathrm{RD},\mathrm{MD})}$ represents today's energy density value of radiation or matter. Specifically, during radiation-dominated (RD) era ($10^{-32} \leq t \leq 10^{12}$): $\rho_{0}^{\mathrm{RD}}\approx 7.86 \times 10^{-31} \frac{kg}{m^{3}}$ denotes the total present-day radiation energy density, including the contribution from standard relativistic neutrinos, while during matter-dominated (MD) epoch ($10^{12} \leq t \leq 10^{17}$): $ \rho_{0}^{\mathrm{MD}}\approx 2.68 \times 10^{-27}  \frac{kg}{m^{3}}$ which includes both the cold dark matter and baryonic components \cite{Aghanim}. Besides, $ G=6.67430 \times 10^{-11} \frac{N m^2}{kg^2}$ is the 2018 CODATA Newtonian constant of gravitation. 
	\

The crossover is dictated by the progression of cosmic time itself. 
The switch occurs precisely because time reaches the specific age $t_{\rm eq}$, 
at which the instantaneous radiation and matter densities satisfy $\rho^{\mathrm{RD}}(t_{\rm eq}) = \rho^{\mathrm{MD}}(t_{\rm eq})$. 
Crucially, because the photon bath cools adiabatically, the scale factor and the photon temperature $T_\gamma(t)$ share a perfect one-to-one correspondence, $a(t) = T_0 / T_\gamma(t)$. 
Consequently, the arrival at $t_{\rm eq}$ is uniquely marked by the thermodynamic threshold $T_\gamma(t_{\rm eq}) = T_{\rm eq} \equiv T_0\left(\frac{\rho_{0}^{\mathrm{MD}}}{\rho_{0}^{\mathrm{RD}}}\right) \simeq 0.8\ {\rm eV},$ which is the single temperature at which the two components contribute equally to the total energy budget.
It is simply the moment when the cosmic stopwatch ticks past $t_{\rm eq}$, as read off the cooling temperature of the universe.
Consequently, this construction guarantees a smooth, self-consistent, and exact analytic connection between the radiation and matter dominated phases, with the transition governed  by the cosmic thermal history via the monotonic time-evolution of the photon temperature.

	\
	
	We now examine the asymptotic behavior of the proposed scale factor across the distinct cosmic epochs: 
	\begin{itemize}
		\item \textbf{Inflation}:  
		In this era \(H_{\mathrm{inf}}=10^{37}\,\mathrm{s}^{-1}\approx 6.6\times10^{12}\,\mathrm{GeV}\), and we require \(k_{\mathrm{inf}}\ge10^{39}\,\mathrm{s}^{-1}\) so that \(k_{\mathrm{inf}}t_{\mathrm{inf}}\gg1\).
		This leads to $ a_{\mathrm{inf}} (t)  \approx e^{H_{\mathrm{inf}} t}$, representing the initial exponential expansion of the inflation. 
		
		\item \textbf{Intermediate era}:  
			 During this epoch $(kt)_{\mathrm{inter}}\ll1$, then according to each era the corresponding scale factor can be generated, consistent with the standard Friedmann prediction, as follows: 
			 
		 Radiation-domination: using \(1-e^{-k t}\approx k t\) and \(b\approx1/2\), we obtain \(a\approx (k_{\mathrm{RD}}t)^{1/2}\). The Hubble parameter from Eq.~\eqref{eq:Hubble_mod} below gives \(\mathcal{H}\approx 1/(2t)\), matching the standard radiation-era expansion.
		 
		 Matter-domination (\(t\gg t_{\mathrm{eq}}\), \(k_{\mathrm{MD}}t\ll1\), \(b\approx2/3\)): \(a\approx (k_{\mathrm{MD}}t)^{2/3}\) and \(\mathcal{H}\approx 2/(3t)\).

		\item \textbf{Dark energy era} (\(t\gtrsim10^{17}\,\mathrm{s}\)): In this regime 
		\(H_{\mathrm{DE}}=2.2\times10^{-18}\,\mathrm{s}^{-1}\approx1.45\times10^{-42}\,\mathrm{GeV}\), so at late times (\(t\to\infty\)) \(a\approx e^{H_{\mathrm{DE}}t}\), corresponding to an asymptotic de Sitter state.
	\end{itemize}
Thus, the single expression \eqref{eq:scale} seamlessly recovers all four standard expansion regimes. 
	
	\section{Smooth transitions via sigmoid functions}
	\label{sec:smooth}
	
	The piecewise definitions are convenient for physical insight but lead to discontinuities at the transition times. To obtain a globally smooth and differentiable scale factor we employ sigmoid function~\cite{Danca2015, Martinelli2023}: 	\[
	\sigma(t;t_T,\Delta_T)=\frac{1}{1+\exp\!\bigl(-\frac{t-t_T}{\Delta_T}\bigr)} .
	\] 	where the parameters \(t_T\) and \(\Delta_T\) control the transition time and its duration. With physically motivated transition timescales and a consistent number of e-folds, we define three such functions for the three transitions as follows:
	\begin{align}
		\sigma_{\mathrm{inf}}(t) &= \frac{1}{1+\exp\!\bigl(-\frac{t-t_{\mathrm{inf}}}{\Delta_{\mathrm{inf}}}\bigr)}, &
		t_{\mathrm{inf}} &\sim 10^{-32}\,\mathrm{s}, & \Delta_{\mathrm{inf}} &\sim 10^{-36}\,\mathrm{s}, \nonumber\\
		\sigma_{\mathrm{eq}}(t) &= \frac{1}{1+\exp\!\bigl(-\frac{t-t_{\mathrm{eq}}}{\Delta_{\mathrm{eq}}}\bigr)}, &
		t_{\mathrm{eq}} &\sim 10^{12}\,\mathrm{s}, & \Delta_{\mathrm{eq}} &\sim 10^{11}\,\mathrm{s}, \label{eq:sigmoids}\\
		\sigma_{\mathrm{DE}}(t) &= \frac{1}{1+\exp\!\bigl(-\frac{t-t_{\mathrm{DE}}}{\Delta_{\mathrm{DE}}}\bigr)}, &
		t_{\mathrm{DE}} &\sim 10^{17}\,\mathrm{s}, & \Delta_{\mathrm{DE}} &\sim 10^{16}\,\mathrm{s}. \nonumber
	\end{align}
	
	 Using these sigmoids we construct the continuous functions \(H(t)\) and \(k(t)\): 	
	\begin{align}
		H(t) &= H_{\mathrm{inf}} \bigl[1 - \sigma_{\mathrm{inf-RD}}(t)\bigr] 
		+ \sigma_{\mathrm{inf-RD}}(t) \biggl\{ 
		\bigl[1 - \sigma_{\mathrm{RD-MD}}(t)\bigr] H_{\mathrm{RD}}(t) \notag \\
		&\quad + \sigma_{\mathrm{RD-MD}}(t) \Big[ \bigl(1 - \sigma_{\mathrm{DE}}(t)\bigr) H_{\mathrm{MD}}(t) 	+ H_{\mathrm{DE}}(t) \sigma_{\mathrm{DE}}(t) \Big]
		\biggr\},
		\label{eq:H}
	\end{align}
which serves as a structural interpolation function controlling the early time inflationary phase and late time acceleration through exponential behavior, and 
	\begin{align}
		k(t) &= k_{\mathrm{inf}} \bigl[1 - \sigma_{\mathrm{inf-RD}}(t)\bigr] 
		+ k_{\mathrm{RD}}\, \sigma_{\mathrm{inf-RD}}(t) \bigl[1 - \sigma_{\mathrm{RD-MD}}(t)\bigr] \notag \\
		&\quad + k_{\mathrm{MD}}\, \sigma_{\mathrm{RD-MD}}(t) \bigl[1 - \sigma_{\mathrm{MD-DE}}(t)\bigr] 
		+ k_{\mathrm{DM}}\, \sigma_{\mathrm{MD-DE}}(t).
		\label{eq:k}
	\end{align}
	as a key parameter governing the energy injection or dissipation, depending on the era.
	  
\begin{figure}[H]
		\centering
		\includegraphics[width=0.58\linewidth]{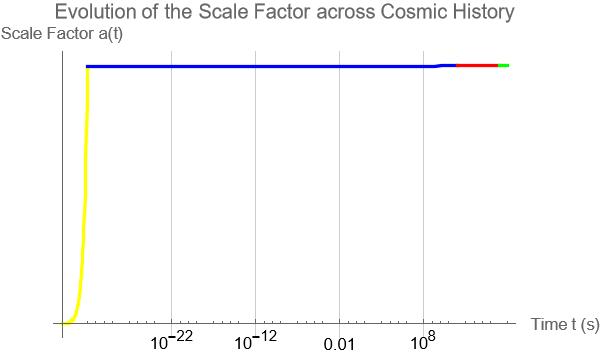}
		\caption{Global view of \(a(t)\) from \(10^{-36}\,\mathrm{s}\) to the present on a log-log scale. Inflation appears as a sharp exponential rise, while the post-inflationary epochs are compressed and appear flattened. This is a visual artifact of the logarithmic time axis, which compresses cosmic history, and does not indicate suppressed expansion. }
		\label{fig:final}
	\end{figure}

\begin{figure}[H]
		\centering
		\includegraphics[width=0.45\textwidth]{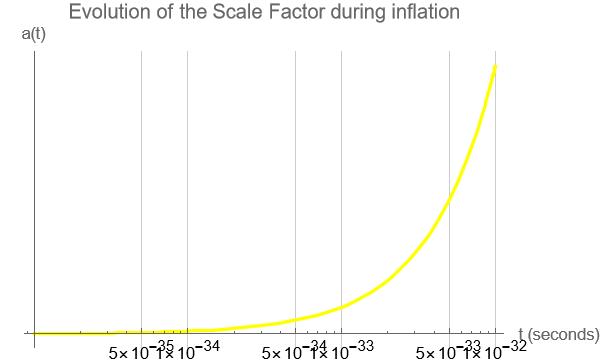}
		\hfill
		\includegraphics[width=0.45\textwidth]{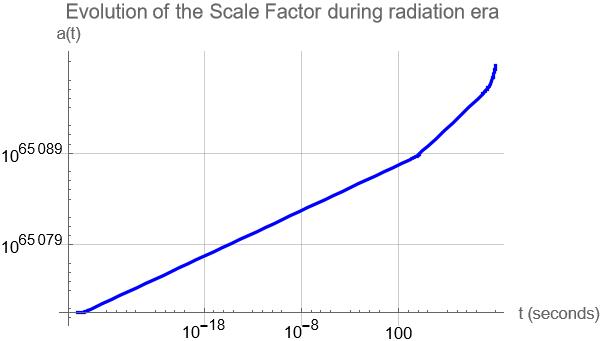}
		
		\vskip\baselineskip
		
		\includegraphics[width=0.45\textwidth]{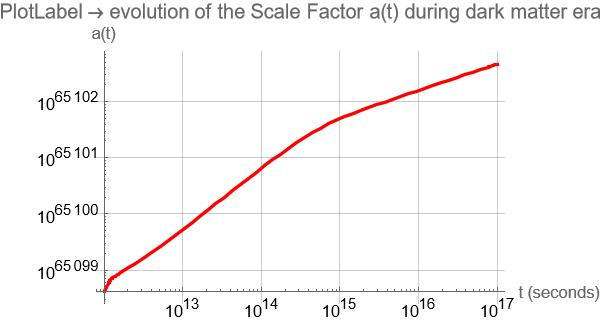}
		\hfill
		\includegraphics[width=0.45\textwidth]{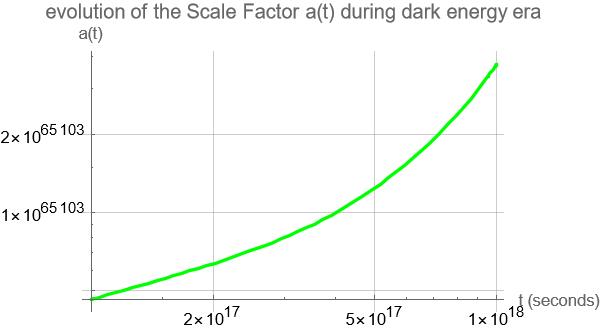}
		
		\caption{ Zoomed views of the scale factor evolution in each epoch separately }
		\label{fig:four_figures}
	\end{figure}

	These parameters are everywhere continuous and differentiable.
		The scale factor itself is then obtained by numerically integrating \(\dot{a}/a \) and with the initial condition chosen so that \(a(t_0)=1\) at the present time. Figures~\ref{fig:final} and~\ref{fig:four_figures} display the scale factor across all epochs, confirming a smooth and monotonic expansion that matches the expected slopes in each era. 
		Figure ~\ref{fig:final} shows the global evolution of the scale factor across cosmological epochs, with zoomed-in plots in Figure ~\ref{fig:four_figures}. Colors indicate each era in both figures.
		
		It should be noted that on the log-log plot, the exponential expansion during inflation manifests as a distinct, steep rise, while the post-inflationary epochs appear comparatively flattened. This apparent flattening, however, is primarily a visual artifact of the logarithmic time compression, which maps the vast temporal interval from reheating to the present onto a narrow window, rather than suppressed cosmic growth. The rapid expansion following inflation is simply rendered smooth and shallow at this global scale.  
		
Therefore, this modified scale factor offers a unified and continuous global formulation of cosmic expansion, replacing piecewise methods and modeling the expansion history in agreement with SNIa, CMB, and BAO observations \cite{Aghanim}, \cite{Eisenstein}.

	\section{Brane-world interpretation and effective variable tension}
	\label{sec:brane}
	
	This model establishes a direct correspondence between the scale-factor dynamics and a time-dependent brane tension, thereby offering a compelling realization within the brane-world paradigm. While the construction is phenomenological in origin, this mapping is precise: 
	identifying the scale-modulation parameter with the ratio \(x(t)\sim\rho(t)/\lambda(t)\) directly links the effective cosmic evolution to a dynamical brane tension. 
   	\subsection{Hubble Parameter}
   	To make this correspondence explicit, we derive the Hubble parameter of the model, denoted by $\mathcal{H}(t)$. Equation \eqref{eq:scale} yields: 
   		\begin{equation}
		\mathcal{H}(t)\equiv\frac{\dot{a}}{a}= H(t)+\dot{H}(t)t +\dot{b}(t)\ln\ \bigl(1-e^{-k(t)t}\bigr)
		+ b(t)\frac{e^{-k(t)t}}{1-e^{-k(t)t}}\bigl[k(t)+t\dot{k}(t)\bigr]. \label{eq:Hubble_mod}
	\end{equation}
	This time-dependent Hubble parameter represents a significant modification to the standard cosmology. The terms $H(t) + \dot{H}(t)t$ capture both the background expansion rate and its evolution, improving the description of the universe's expansion and increasing sensitivity to deviations from standard expansion histories.
	The remaining terms introduce novel corrections to the modified Hubble parameter. They contain the factors $\ln(1-e^{-x})$ and $1/(e^{x}-1)$ with $x\equiv k(t)t$, which have the same mathematical forms as the bosonic grand-potential contribution and the Bose–Einstein occupation factor, respectively \cite{Vakili}. Whether this analogy reflects an actual physical connection to thermal field theory or quantum effects is an open question lies beyond the scope of present work and we leave it for future investigation.
	
	Recent studies demonstrate that two parametrizations of the Hubble parameter, the power-law and logarithmic corrections, provide a convincing fit to current observational data. These corrections effectively model observed cosmic acceleration and refine the standard 
	$\Lambda$CDM model \cite{Koussour}. 
	The present model naturally generates such corrections.
	
	Crucially, the dimensionless combination \(x(t)\) provides a control parameter that continuously selecting the appropriate expansion regime at each cosmic epoch: 
In the very early universe (at high energies), where $x(t)\gg 1$, the second exponential term vanishes, leading to $	{\mathcal H}(t) \approx H_{\mathrm{inf}}$.  
During the universe's intermediate expansion era $\bigl(x(t)\ll 1\bigr)$, the effective Hubble parameter can be approximated as  $ {\mathcal H(t)} \approx H(t) + \dot{H}(t) t + \dot{b}(t) \ln(x(t)) +\frac{b(t)}{t}\left[1+t\eta(t) - x(t)-x(t)t \eta(t) \right]$, 
with $\eta(t)\equiv\frac{ \dot{k}(t)}{k(t)}$. Since other terms are subdominant compared to $\frac{b(t)}{t}$ during power-law expansion, then, ${\mathcal H}(t)\sim \frac{b(t)}{t}$. This means, the standard linear term in the Friedmann equation dominates, leading to radiation and matter-dominated epochs characterized by the energy density dependence $\rho_{eff}(t)=\frac{3{\mathcal H}^{2}(t)}{8\pi G}\approx \frac{3b^{2}(t)}{8\pi G t^2}$. 
At late times where $x(t)\gg 1$, this model enters a phase of accelerated expansion again, during which the Hubble parameter approaches $\mathcal H(t)\approx H_{DE}$, with an effective energy density tends to a constant $\rho_{eff}(t)=\dfrac{3H_{DE}^{2}}{8\pi G}$. 
These asymptotic behaviors demonstrate that $x(t)$ naturally governs the cosmic evolution in this model.
 
		\subsection{Brane-world Cosmology}
	To interpret $x(t)$ physically, we recall the essential features of the Randall-Sundrum type-II (RSII) brane-world setup, where the effective Friedmann equation on a flat 3-brane is~\cite{Randall1999a, Randall1999b}
	\begin{equation}
		\mathcal{H}_{\mathrm{RS}}^2 = \frac{8\pi G}{3}\rho\left(1+\frac{\rho}{2\lambda}\right)+\frac{\Lambda_4}{3}+\frac{\mathcal{C}}{a^4},
	\end{equation}
	In this relation, \(\lambda\) is the constant brane tension, \(\rho\) the total energy density,
	\(\Lambda_4\) an effective cosmological constant and \(\mathcal{C}\) a dark
	radiation term. A central quantity is the ratio $\frac{\rho}{\lambda}$ which controls the magnitude of the quadratic correction to the Friedmann equation. 
	 At high energies (\(\rho\gg\lambda\)) the \(\rho^2\) term dominates, modifying the expansion. In this regime using the brane tension $\lambda = (1-10)\times 10^{57}\text{ GeV}^{4}$ derived from CMB normalisation~\cite{Mounzi2017}, one finds $\rho/\lambda \sim 10^{3}-10^{4}$. After inflation, the energy density drops by many orders of magnitude. During Big Bang Nucleosynthesis (BBN) and throughout the subsequent radiation and matter-dominated eras, \(\rho/\lambda \ll 1\); thus the linear term of the Friedmann equation prevails and standard cosmology is recovered with high precision~\cite{Langlois2003}. At late times, the dark-energy density \(\rho_{\mathrm{DE}} \sim 10^{-47}\) GeV$^4$ also yields \(\rho/\lambda \ll 1\). Table~\ref{tab:rholambda_RSII} summarises the \(\rho/\lambda\) classification for each era in this model.
		
	\begin{table}[htbp]
		\centering
		\caption{The $\rho/\lambda$ Ratio Across Cosmic Eras in Constant Brane Tension (RSII) Model}
		\begin{tabular}{lcccc}
			\toprule
			Cosmic Era & $\rho$ [GeV$^4$] & $\lambda$ [GeV$^4$] & $\rho/\lambda$ & Dominant Regime \\
			\midrule
			Inflation & $\sim 10^{60}$ & $(1-10) \times 10^{57}$ & $\gg1$ & High-energy  \\
			BBN & $\sim 10^{-17}$ & $\gtrsim 10^{57}$ & $\ll 1$ & Low-energy  \\
			Matter-Radiation Equality & $\lesssim 10^{-37}$ & $\gtrsim 10^{57}$ & $\ll 1$ & Low-energy \\
			Late time (Dark Energy) & $\sim 10^{-47}$ & $\gtrsim 10^{57}$ & $\ll 1$ & Low-energy  \\
			\bottomrule
		\end{tabular}
		\label{tab:rholambda_RSII}
	\end{table}
	
    Accordingly, in the standard RSII scenario with constant brane tension, the energy density $\rho(t)$ decreases monotonically with cosmic expansion, so the ratio $\rho(t)/\lambda$ remains small throughout all post-nucleosynthesis epochs. Consequently, such constant-tension branes do not produce late-time accelerated expansion unless an additional dark-energy component (e.g., a cosmological constant) is introduced by hand~\cite{Dimopoulos2002}.

	A compelling alternative is the chrono-brane model proposed by Garc\'ia-Aspeitia et al.~\cite{GarciaAspeitia2018a}, in which the brane tension is promoted to a dynamic degree of freedom that varies with redshift as $\lambda(z) = \lambda_0(1+z)^n$. Here $\lambda_0$ is the present day tension and $n$ a constant exponent. The model is built upon the RSII braneworld, but with a crucial twist. For the observationally preferred value $n\simeq6.19$, the brane tension increases rapidly toward higher redshift. In the radiation-dominated regime, where $\rho\propto(1+z)^4$, the density-to-tension ratio therefore scales as $		\frac{\rho}{\lambda}\propto(1+z)^{4-n}
		\simeq(1+z)^{-2.19},$
	whereas during matter domination, $		\frac{\rho}{\lambda}\propto(1+z)^{3-n}
		\simeq(1+z)^{-3.19}.$
For the best-fit parameters, $\rho/\lambda$ is  $\simeq3.26\times10^{-22}$ at BBN ($z\sim3\times10^8$, $t\sim10^2{\rm s}$), $\simeq4.62\times10^{-11}$ at $z\simeq3400$ ($t\sim10^{12}{\rm s}$), and increases to $\simeq4.44$ at the present epoch. See Appendix ~\ref{app:derivation}  for the detailed derivation. Thus, the cosmological evolution is characterized by an extremely strong hierarchy $\rho\ll\lambda$ at early times, which progressively weakens toward late times until $\rho>\lambda$ today. The ratio $\rho/\lambda$ consequently provides a useful measure of the evolving density-to-tension hierarchy, rather than the exact fractional correction to the Friedmann equation.

	It should be noted that in this model, all results apply strictly to the post-inflationary universe. Any statement about behaviour at high redshift in this paper refers to redshifts $z\lesssim10^4$ (nucleosynthesis and later), not to the GUT‑scale redshifts of inflation ($z\sim10^{25}$). 	Table~\ref{tab:chrono-brane} lists approximate values of \(\rho/\lambda\) for each era in this model.

		\begin{table}[htbp]
			\centering
			\caption{physically consistent cosmological parameters for the chrono-brane model.}
			\begin{tabular}{lcccc}
				\toprule
				Cosmic Era &  $\rho$ [GeV$^4$] & $\lambda$ [GeV$^4$] & $\rho/\lambda$ & Dominant Regime \\
				\midrule
				BBN  & $\sim10^{-17}$ & $8.35\times10^{4}$ & $\sim10^{-22}$ & Linear Term (GR) \\
				Matter-Radiation Equality & $\sim10^{-37}$ & $2.04\times 10^{-26}$ & $\sim 10^{-11}$ & Linear Term (GR) \\
				Present Epoch & $\sim10^{-47}$ & $2.81 \times 10^{-48}$ & $\sim4$ & Quadratic Term \\
				\bottomrule
			\end{tabular}
				\label{tab:chrono-brane}
		\end{table}

	\subsection{Phenomenological Interpretation}

	The chrono-brane model is formulated primarily to describe the late-time cosmological dynamics rather than inflation or the very early Universe.
	At high redshifts, corresponding to the epochs of BBN and matter–radiation equality, the density to tension ratio remains very small, $\rho/\lambda\ll1$, reflecting a strong hierarchy between the cosmological energy density and the brane tension. In this regime, the quadratic brane correction is strongly suppressed, and the background evolution approaches the standard radiation and matter-dominated behavior. As the Universe expands toward lower redshifts, $\rho/\lambda$ increases progressively, eventually approaching values of order unity at late times. This evolution marks the transition from a regime in which brane effects are negligible to one in which the brane sector can become dynamically relevant to the background expansion. In the present framework, the evolution of the brane tension with the cosmic expansion compensates the dilution of the cosmological energy density in such a way that the associated quadratic brane correction remains constant. The resulting chrono-brane contribution can therefore provide a mechanism for late-time accelerated expansion without introducing a conventional cosmological constant.

	This behaviour is qualitatively analogous to that of our parameter $x(t)$, which captures the same relative evolution as the ratio \(\rho/\lambda\), with the important distinction that in our construction, all cosmic epochs, including inflation, emerge intrinsically from the model structure, with no need for an externally imposed dark-energy component:	
    During inflation, $x(t) \sim \rho/\lambda \gg 1$, and the Hubble parameter is approximately constant, $H \simeq H_{\mathrm{inf}}$. Matching this to the high-energy limit of the brane Friedmann equation, 
    $H^2 \simeq (4\pi G/3\lambda)\rho^2$, yields an effective energy density $\rho_{\mathrm{eff}} \sim \sqrt{\lambda} H_{\mathrm{inf}}$, with $H_{\mathrm{inf}} \simeq 6.6 \times 10^{12} \, \mathrm{GeV}$. 
    This correspondence places the inflationary dynamics precisely in the high-energy regime of the RSII scenario.
	During the intermediate epochs where \(x(t)\sim \rho/\lambda\ll1\) we have \(\mathcal{H}\sim b(t)/t\) and the energy density scales as \(\rho\sim 3\mathcal{H}^2/(8\pi G)\), effectively transitions to 4D general relativity behavior.
	At late times, \(x(t)\sim \rho/\lambda\) again becomes large, and a small constant \(H_{\mathrm{DE}}\) can be interpreted as a low-energy effective cosmological constant on the brane with the effective energy density $	\rho_{\mathrm{eff}} = \frac{3H_{\mathrm{DE}}^2}{8\pi G_4} \approx 3.1\times10^{-47}~\text{GeV}^4$, achieving late-time accelerated expansion.
	Consequently, parameter $x(t)$ effectively captures the behavior of both paradigms, mimicking the RSII inflationary dynamics and the chrono-brane late-time limit, within a single, unified phenomenological framework. While the construction is effective rather than derived from a fundamental action, it demonstrates that a single, continuous parameter can reproduce the observational successes of both pictures across all epochs. The model thereby offers a valuable unified perspective that is both analytically tractable and observationally testable, without invoking abrupt phase transitions or a cosmological constant.

	If we postulate the correspondence $
		x(t)=k(t)t \sim \frac{\rho(t)}{\lambda(t)}, \label{eq:kt_lambda}$
	then the factor \((1-e^{-k(t)t})^{b(t)}\) naturally suppresses the brane corrections when the density is low relative to the tension. 
		Using the scale factor \eqref{eq:scale} an explicit expression for \(\lambda(t)\) can be  obtained as:
		\begin{equation}
		\lambda(t) \sim \frac{\rho(t)}{k(t)t} = -\frac{\rho(t)}{\ln\!\left[1-\left(\frac{a(t)}{e^{H(t)t}}\right)^{1/b(t)}\right]}. \label{eq:lambda}
	\end{equation}
	This relation shows that the entire cosmic history can be viewed as the evolution of a brane whose tension \(\lambda(t)\) varies with time. This interpretation unifies all epochs under a single framework without invoking a separate cosmological constant. We emphasise that \(\rho(t)\) in this equation is the effective energy density that would appear in the standard Friedmann equation, not necessarily the true brane energy density; the mapping is phenomenological but provides a clear physical picture.

Therefore, the unified scale factor ansatz in this work finds a compelling realization as a brane-inspired effective framework. The required time-varying tension emerges naturally in variable-tension models and consistently describes the full cosmic history. While not derived from first principles, this interpretation grounds the phenomenological construction in the brane world paradigm via the mapping $k(t)t \sim \frac{\rho(t)}{\lambda(t)}, \label{eq:kt_lambda}$ opening avenues for future investigation.

	\section{Discussion and future directions}
	\label{sec:Discussion}

 The main aim of this work is not to challenge the $\Lambda$CDM model, but to establish the observational viability of a unified phenomenological ansatz for the scale factor across the entire cosmic timeline, while providing a brane-inspired effective description of the cosmic expansion.
Given the phenomenological nature of the framework, two important features emerge that deserve further theoretical investigation:

\paragraph{A dynamical perspective on cosmic acceleration:}

Unlike the standard $\Lambda$CDM paradigm, which posits a fixed, non-dynamical cosmological constant decoupled from the matter and radiation sectors, our framework encodes the transition to late-time acceleration intrinsically within the functional form of the scale factor $a(t)$. Acceleration therefore emerges not as an isolated late-time add-on, but as an integrated feature of the same dynamical architecture that dictates all prior cosmic epochs. 
This construction naturally bridges the evolutionary phases, obviating both the need for abrupt phase transitions and the requirement of a strictly constant $\Lambda$ term.

\paragraph{A speculative hint of deeper structure:}

The model bridges inflation and late-time acceleration through the asymptotic limits of the single dimensionless parameter $x(t)$. This structural unification raises the possibility of a hidden symmetry or conserved quantity underlying the cosmic evolution. Whether this correspondence reflects an exact symmetry, a holographic duality, or simply an emergent feature of the effective dynamics remains an open question.
This framework does not attempt to derive such a symmetry from the underlying theory; rather, it serves as a heuristic guide. Within this limited but observationally testable scope, the model offers a modest step toward integrating the cosmic evolution into a more unified and coherent framework.

Looking ahead, our phenomenological parametrization can serve as a useful benchmark for comparison and a flexible framework for hypothesis testing. Key direction for future work include investigating whether the adopted scale-factor evolution can be derived from a physically motivated action, for example within $f(R)$ gravity. Until such a theoretical foundation is established, the model is best regarded as a flexible and observationally consistent effective framework for interpreting the expansion history, providing a complementary perspective on the standard cosmological paradigm.

	\section{Conclusions}
	\label{sec:conclusion}

	We present a single analytic scale factor \(a(t)=e^{H(t)t}\bigl(1-e^{-k(t)t}\bigr)^{b(t)}\) that unifies the cosmic expansion history from inflation to the present in a smooth and differentiable manner. The model eliminates the need for piecewise definitions, providing a compact representation of cosmic evolution in agreement with SNIa, CMB, and BAO observations.
	 
  The key dimensionless variable $x(t)=k(t)t$ of this work, governs the entire asymptotic history, smoothly connecting inflation, the radiation and matter-dominated intermediate phases, and the late-time accelerating epoch within a single dynamical framework. 
Central to this construction is the identification of $x(t)\sim\rho(t)/\lambda(t)$ of the total energy density to a time-dependent brane tension. This interpretation naturally accommodates a variable tension, demonstrating that a constant brane tension is incompatible with the full cosmic history and thereby motivating a dynamical $\lambda(t)$. 
This construction unifies the observational successes of the RSII and chrono-brane paradigms across all epochs. Although phenomenological in origin, it provides a compelling, analytically tractable alternative to $\Lambda$CDM, requiring neither a fixed cosmological constant nor abrupt phase transitions.
The reconstruction of $\lambda(t)$ from the scale factor $a(t)$ then encodes the entire expansion history as a single brane trajectory governed by the evolving tension. This establishes a coherent effective framework that unifies the early and late-time accelerated phases within a single, continuous bridge to brane-world cosmology.

	This work demonstrates that a phenomenologically motivated scale factor can be more than a mere fitting function; it can encode a rich physical interpretation.
	It suggests the onset of cosmic acceleration could be intrinsically via the form of scale factor, imposed by its underlying theory. Thus the acceleration emerges naturally from the same unified structure that governs earlier cosmological epochs. 
	The model is observationally viable (matching supernova, CMB, and BAO data, as shown in the figures) and opens several avenues for future research: 
	(i) deriving the ansatz from a fundamental action or from a specific brane‑world setup with a dynamical tension, 
	(ii) testing the model’s predictions for the cosmic microwave background power spectrum and large‑scale structure, and 
	(iii) exploring the quantum‑statistical analogy more rigorously, possibly linking to Bose-Einstein condensation of dark matter or dark energy.

\appendix

\section{${\frac{\rho}{\lambda}}$ Hierarchy at BBN, Matter-Radiation Equality, and Late Times}
\label{app:derivation}

In this appendix, we quantify the hierarchy between the cosmological energy density and the dynamical brane tension of the chrono brane model at three characteristic epochs: Big Bang nucleosynthesis (BBN), matter-radiation equality, and the present epoch. This comparison provides a direct measure of the importance of the brane correction and is particularly useful for
assessing the recovery of the standard cosmological evolution at early times.

For the joint observational constraints considered in \cite{GarciaAspeitia2018a}, the
best-fit parameters are
\begin{equation}
	h=0.706,\qquad
	\Omega_{m0}=0.31,\qquad
	n=6.19,\qquad
	\lambda_0=2.81\times10^{-12}\ {\rm eV}^4.
	\label{eq:app_bestfit}
\end{equation}
The redshift-dependent brane tension is parametrized as
\begin{equation}
	\lambda(z)=\lambda_0(1+z)^n.
	\label{eq:app_lambda_z}
\end{equation}

The present critical density is
\begin{equation}
	\rho_{c0}
	=
	8.070\times10^{-11}h^2\ {\rm eV}^4,
\end{equation}
which, for $h=0.706$, gives
\begin{equation}
	\rho_{c0}
	=
	4.02238\times10^{-11}\ {\rm eV}^4
	=
	4.02238\times10^{-47}\ {\rm GeV}^4.
	\label{eq:app_rhoc}
\end{equation}
Throughout this appendix we use
\begin{equation}
	1\ {\rm eV}^4=10^{-36}\ {\rm GeV}^4.
	\label{eq:app_units}
\end{equation}

The radiation density parameter is obtained from
\begin{equation}
	\Omega_{r0}
	=
	2.469\times10^{-5}h^{-2}
	\left(1+0.2271N_{\rm eff}\right),
	\label{eq:app_Omega_r}
\end{equation}
with $N_{\rm eff}=3.04$. Hence,
\begin{equation}
	\Omega_{r0}
	=
	8.37331\times10^{-5}.
	\label{eq:app_Omega_r_num}
\end{equation}

The matter and radiation energy densities therefore evolve according to
\begin{align}
	\rho_m(z)
	&=
	\rho_{c0}\Omega_{m0}(1+z)^3,
	\label{eq:app_rhom}\\
	\rho_r(z)
	&=
	\rho_{c0}\Omega_{r0}(1+z)^4.
	\label{eq:app_rhor}
\end{align}
For the purpose of comparing the cosmological matter-radiation sector
with the brane tension, we define
\begin{equation}
	\rho(z)\equiv \rho_m(z)+\rho_r(z).
	\label{eq:app_rho_total}
\end{equation}
This quantity should not be confused with the present critical density,
which also encodes the effective late-time contribution responsible for
the accelerated expansion.

\subsection{BBN epoch}

The BBN epoch is taken to correspond to
\begin{equation}
	z_{\rm BBN}\simeq3\times10^8.
	\label{eq:app_z_bbn}
\end{equation}
At this epoch, radiation dominates strongly, and Eq.~\eqref{eq:app_rhor}
gives
\begin{equation}
	\rho_r(z_{\rm BBN})
	=
	\rho_{c0}\Omega_{r0}(1+z_{\rm BBN})^4.
\end{equation}
Substitution of the numerical values yields
\begin{equation}
	\rho_{\rm BBN}
	\simeq
	2.72816\times10^{19}\ {\rm eV}^4
	=
	2.72816\times10^{-17}\ {\rm GeV}^4.
	\label{eq:app_rho_bbn}
\end{equation}

The corresponding brane tension follows directly from
Eq.~\eqref{eq:app_lambda_z}:
\begin{align}
	\lambda_{\rm BBN}
	&=
	2.81\times10^{-12}
	(3\times10^8)^{6.19}\ {\rm eV}^4
	\nonumber\\
	&=
	8.35770\times10^{40}\ {\rm eV}^4
	\nonumber\\
	&=
	8.35770\times10^{4}\ {\rm GeV}^4.
	\label{eq:app_lambda_bbn}
\end{align}
Consequently, the dimensionless density-to-tension ratio is
\begin{equation}
	\left.\frac{\rho}{\lambda}\right|_{\rm BBN}
	=
	\frac{2.72816\times10^{-17}}
	{8.35770\times10^{4}}
	=
	3.26425\times10^{-22}.
	\label{eq:app_ratio_bbn}
\end{equation}
Thus,
\begin{equation}
	\boxed{
		\left.\frac{\rho}{\lambda}\right|_{\rm BBN}
		\simeq3.26\times10^{-22}
	}.
\end{equation}
The extremely small value demonstrates that the cosmological energy
density is many orders of magnitude below the brane tension during BBN.

\subsection{Matter-radiation equality}

For comparison with the conventional cosmological matter--radiation
equality epoch, we adopt
\begin{equation}
	z_{\rm eq}\simeq3400.
	\label{eq:app_z_eq}
\end{equation}
At this redshift, the matter density is
\begin{align}
	\rho_m(z_{\rm eq})
	&=
	\rho_{c0}\Omega_{m0}(1+z_{\rm eq})^3
	\nonumber\\
	&=
	4.90529\times10^{-37}\ {\rm GeV}^4,
	\label{eq:app_rhom_eq}
\end{align}
while the radiation density is
\begin{align}
	\rho_r(z_{\rm eq})
	&=
	\rho_{c0}\Omega_{r0}(1+z_{\rm eq})^4
	\nonumber\\
	&=
	4.50616\times10^{-37}\ {\rm GeV}^4.
	\label{eq:app_rhor_eq}
\end{align}
Thus, evaluating the fitted cosmological densities at $z=3400$ gives
\begin{equation}
	\rho(z_{\rm eq})
	=
	\rho_m(z_{\rm eq})+\rho_r(z_{\rm eq})
	=
	9.41145\times10^{-37}\ {\rm GeV}^4.
	\label{eq:app_rho_eq}
\end{equation}

The brane tension at the same redshift is
\begin{align}
	\lambda_{\rm eq}
	&=
	\lambda_0(1+z_{\rm eq})^n
	\nonumber\\
	&=
	2.81\times10^{-12}(3401)^{6.19}\ {\rm eV}^4
	\nonumber\\
	&=
	2.03869\times10^{10}\ {\rm eV}^4
	\nonumber\\
	&=
	2.03869\times10^{-26}\ {\rm GeV}^4.
	\label{eq:app_lambda_eq}
\end{align}
It follows that
\begin{align}
	\left.\frac{\rho}{\lambda}\right|_{\rm eq}
	&=
	\frac{9.41145\times10^{-37}}
	{2.03869\times10^{-26}}
	\nonumber\\
	&=
	4.61642\times10^{-11}.
	\label{eq:app_ratio_eq}
\end{align}
Therefore,
\begin{equation}
	\boxed{
		\left.\frac{\rho}{\lambda}\right|_{\rm eq}
		\simeq4.62\times10^{-11}
	}.
\end{equation}

It is worth emphasizing that $z_{\rm eq}=3400$ is the conventional
equality redshift adopted here. If equality is instead defined strictly
from the fitted present-day densities through
$\rho_m(z_{\rm eq})=\rho_r(z_{\rm eq})$, one obtains
\begin{equation}
	1+z_{\rm eq}
	=
	\frac{\Omega_{m0}}{\Omega_{r0}}
	\simeq3702,
\end{equation}
corresponding to $z_{\rm eq}\simeq3701$. The resulting
$\rho/\lambda$ is approximately $3.7\times10^{-11}$. The value quoted
in Eq.~\eqref{eq:app_ratio_eq} corresponds specifically to the
conventional choice $z_{\rm eq}=3400$.

\subsection{Late-time universe}

At the present epoch, $z=0$, the brane tension is simply
\begin{equation}
	\lambda_0
	=
	2.81\times10^{-12}\ {\rm eV}^4
	=
	2.81\times10^{-48}\ {\rm GeV}^4.
	\label{eq:app_lambda_0}
\end{equation}

The present matter density is
\begin{equation}
	\rho_{m0}
	=
	\Omega_{m0}\rho_{c0}
	=
	1.24694\times10^{-47}\ {\rm GeV}^4,
\end{equation}
whereas the radiation density is
\begin{equation}
	\rho_{r0}
	=
	\Omega_{r0}\rho_{c0}
	=
	3.36806\times10^{-51}\ {\rm GeV}^4.
\end{equation}
Consequently,
\begin{equation}
	\rho_0
	=
	\rho_{m0}+\rho_{r0}
	=
	1.24727\times10^{-47}\ {\rm GeV}^4.
	\label{eq:app_rho_0}
\end{equation}

The corresponding ratio is therefore
\begin{align}
	\left.\frac{\rho}{\lambda}\right|_0
	&=
	\frac{1.24727\times10^{-47}}
	{2.81\times10^{-48}}
	\nonumber\\
	&=
	4.43870.
	\label{eq:app_ratio_0}
\end{align}
Thus,
\begin{equation}
	\boxed{
		\left.\frac{\rho}{\lambda}\right|_0
		\simeq4.44
	}.
\end{equation}

\subsection{Summary and physical interpretation}

The resulting density-to-tension hierarchy is summarized in
Table~\ref{tab:rho_lambda_hierarchy}.

\begin{table}[htbp]
	\centering
	\caption{Exact cosmological energy density, brane tension, and their ratio at
		BBN, matter-radiation equality, and the present epoch of the chrono brane model }
	\label{tab:rho_lambda_hierarchy}
	\begin{tabular}{lcccc}
		\hline\hline
		Epoch & $z$ & $\rho\,[{\rm GeV}^4]$
		& $\lambda\,[{\rm GeV}^4]$ & $\rho/\lambda$ \\
		\hline
		BBN & $3\times10^8$
		& $2.728\times10^{-17}$
		& $8.358\times10^{4}$
		& $3.264\times10^{-22}$ \\
		
		Matter--radiation equality & $3400$
		& $9.411\times10^{-37}$
		& $2.039\times10^{-26}$
		& $4.616\times10^{-11}$ \\
		
		Present epoch & $0$
		& $1.247\times10^{-47}$
		& $2.810\times10^{-48}$
		& $4.439$ \\
		\hline\hline
	\end{tabular}
\end{table}

The evolution of the ratio can also be understood analytically. During
radiation domination,
\begin{equation}
	\rho\simeq\rho_r\propto(1+z)^4,
\end{equation}
whereas
\begin{equation}
	\lambda\propto(1+z)^n.
\end{equation}
Consequently,
\begin{equation}
	\frac{\rho}{\lambda}
	\propto
	(1+z)^{4-n}
	=
	(1+z)^{-2.19}
\end{equation}
for $n=6.19$. During matter domination,
\begin{equation}
	\frac{\rho}{\lambda}
	\propto
	(1+z)^{3-n}
	=
	(1+z)^{-3.19}.
\end{equation}
Hence, the ratio decreases rapidly toward the early Universe and grows
toward late times. Numerically,
\begin{equation}
	\boxed{
		\frac{\rho}{\lambda}
		\sim
		3.26\times10^{-22}
		\;\longrightarrow\;
		4.62\times10^{-11}
		\;\longrightarrow\;
		4.44
	}
\end{equation}
from BBN through matter-radiation equality to the present epoch.

This hierarchy shows that the brane tension is overwhelmingly larger
than the cosmological matter-radiation density during BBN and around
matter-radiation equality, ensuring that the corresponding
brane correction is strongly suppressed during the early stages of
cosmic evolution. At late times, however, the ratio becomes of order
unity, allowing the brane sector to have a significant impact on the
background expansion and to contribute to the observed late-time
acceleration.

\section*{acknowledgments}
		The author thanks H. Firouzjahi, A. Talebian, and M. H. Namjoo for valuable discussions, and acknowledges the use of computing facilities at IPM.

\section*{Declaration of generative AI and AI-assisted technologies in the manuscript preparation process}

During the preparation of this work, the author(s) used ChatGPT, Claude and DeepSeek to assist with translation, paraphrasing, editing, and improving the language, grammar, clarity, and readability of the manuscript, the primary draft of which was written by the authors. After using this tool/service, the author(s) critically reviewed, verified, and revised all AI-assisted content as needed and take(s) full responsibility for the accuracy, originality, and integrity of the publication.

\end{document}